\documentclass{elsarticle}

\usepackage{fullpage} 

\usepackage{tabularx} 
\newcolumntype{L}[1]{>{\hsize=#1\hsize\raggedright\arraybackslash}X}%
\newcolumntype{R}[1]{>{\hsize=#1\hsize\raggedleft\arraybackslash}X}%
\newcolumntype{C}[1]{>{\hsize=#1\hsize\centering\arraybackslash}X}%

\usepackage{hyperref}

\journal{J. Theor. Biol.; doi:10.1016/j.jtbi.2018.06.002}

\begin{document}

\begin{frontmatter}

\title{Mechanics of epithelial tissue formation}
\author[Delftaddress]{Ruben van Drongelen\corref{equalcontribution}}
\cortext[equalcontribution]{These authors contributed equally.}

\author[Delftaddress,Leidenaddress]{Tania Vazquez-Faci\corref{equalcontribution}}

\author[Delftaddress]{Teun A. P. M. Huijben}

\author[Leidenaddress]{Maurijn van der Zee}

\author[Delftaddress]{Timon Idema
\corref{mycorrespondingauthor}}
\cortext[mycorrespondingauthor]{Corresponding author}
\ead{t.idema@tudelft.nl}

\address[Delftaddress]{Department of Bionanoscience, Kavli Institute of Nanoscience, Delft University of Technology, Delft, The Netherlands}
\address[Leidenaddress]{Institute of Biology, Leiden University, Leiden, The Netherlands}

\begin{abstract}
A key process in the life of any multicellular organism is its development from a single egg into a full grown adult. The first step in this process often consists of forming a tissue layer out of randomly placed cells on the surface of the egg. We present a model for generating such a tissue, based on mechanical interactions between the cells, and find that the resulting cellular pattern corresponds to the Voronoi tessellation of the nuclei of the cells. Experimentally, we obtain the same result in both fruit flies and flour beetles, with a distribution of cell shapes that matches that of the model, without any adjustable parameters. Finally, we show that this pattern is broken when the cells grow at different rates.
\end{abstract}


\end{frontmatter}


\section{Introduction}
Multicellular organisms start life as a single fertilized cell. From this modest beginning, they undergo a developmental process that leads to the formation of complex tissues and organs with a wide range of different functions. Although it has long been appreciated that these various components of an organism have very different mechanical properties, the role of mechanical interactions in the developmental process has only become the focus of detailed studies relatively recently. One of the earliest milestones in this field is the seminal work by Discher et al.~\citep{Discher2005} and Engler et al.~\citep{Engler2006}, who showed that identical stem cells, when placed on substrates of different stiffness, differentiate into cells of tissues with the corresponding stiffness. Cells in living multicellular organisms, however, do not exist on a substrate in isolation; instead, they are part of a tissue that consists of both cells and extracellular material and together form a mechanical system~\citep{Kasza2007}. Moreover, cells react strongly to both direct mechanical interactions with their neighbors~\citep{Vogel2006,Gibson2006,Schwarz2013,Idema2013,Shawky2015,Kaiser2018} and indirect interactions via deformations of a shared substrate~\citep{Tang2011,Majkut2013,Nitsan2016}. Finally, the interior organization of the cell, in particular the position of the nucleus, is also mechanically coupled to its outside environment~\citep{Zemel2015}. To understand how epithelial tissues develop, we thus need a mechanical model coupling the inside to the outside of the cell.

As a model epithelial tissue, we study the first tissue developed in insect embryos, the epithelial blastoderm. This tissue forms as a single layer on top of the yolk. The nuclei of the fertilized egg first divide a couple of times in the egg's interior, then migrate to the surface where they continue to divide, eventually creating a confluent proto-tissue. This proto-tissue is turned into a proper tissue through invagination of the egg's outer (plasma) membrane, which separates the nuclei into cells (cellularization)~\citep{Anderson1972,Foe1983,Handel2000,Mazumdar2002,Lecuit2004,Harris2009,VanderZee2015}. Already during the syncytial stage (i.e., before cellularization), each nucleus is embedded in a full cellular apparatus, including organelles and a cytoskeleton. We present a model for the formation of the epithelial blastoderm. We also study this tissue formation directly in two model organisms: the fruit fly \textit{Drosophila melanogaster} and the flour beetle \textit{Tribolium castaneum}. We find that the touching boundaries of the (proto)cells correspond closely to a Voronoi tessellation of their nuclei, an effect that becomes more pronounced after cellularization. Although Voronoi tessellations have occasionally been used to describe cellular patterns in epithelial tissues~\citep{Honda1978,Sulsky1984,Weliky1990,Sharma2009,Bock2010,Sanchez-Gutierrez2016,Kim2016}, to the best of our knowledge, the fact that the nuclei are located at the centers of the corresponding Voronoi cells has not been shown previously. Tessellations have also been used as a basis for mechanical modeling of cellular tissues, especially in vertex models where forces act on the vertices of a lattice~\citep{Weliky1990,Farhadifar2007,Staple2010,Fletcher2014,Okuda2015,Barton2017,Alt2017,Lin2017}. In contrast, our model faithfully reproduces the Voronoi tessellation, and matches the experimental data quantitatively on a number of geometric and topological measures, irrespective of the choice of the mechanical parameters of the model. We conclude that the mechanical interactions between the (proto)cells in early embryonic epithelial tissues are directly responsible for the observed geometrical cellular patterns of those tissues.

\section{Materials and methods}
\subsection{Model}
\begin{figure}
\begin{center}
\includegraphics{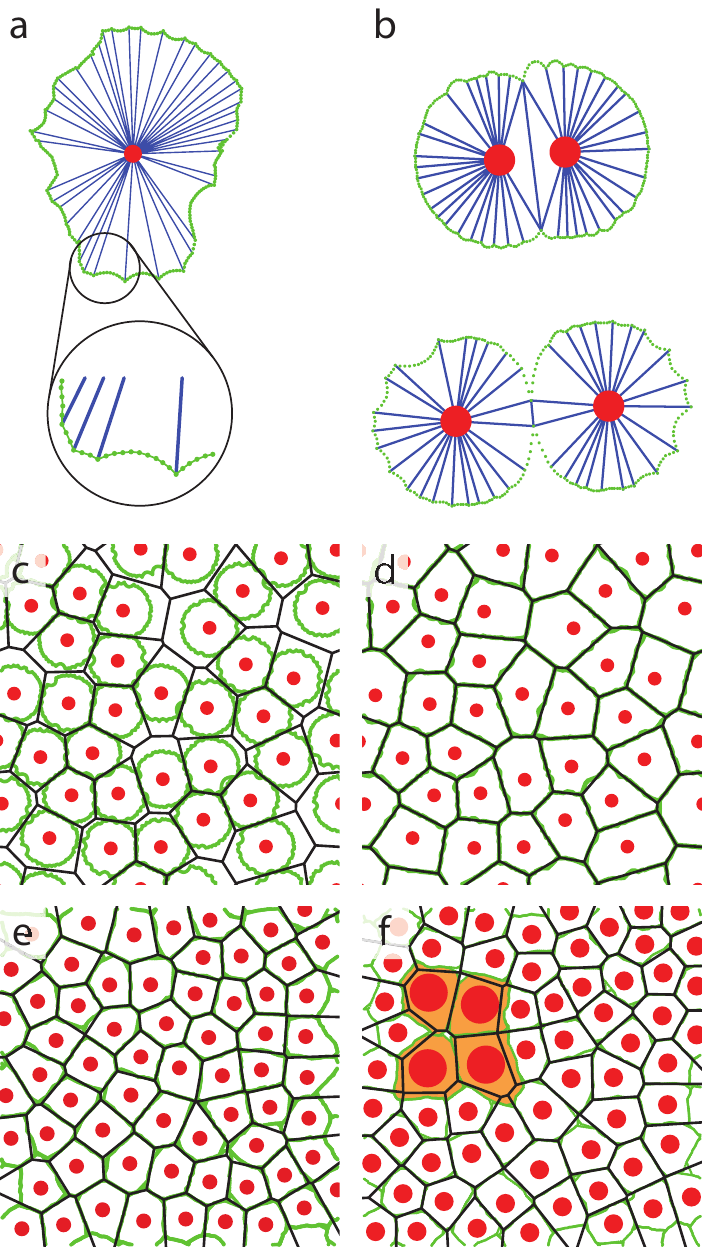}
\end{center}
\caption{Mechanical cell model and simulation results. (a) Cells consist of a sphere representing the nucleus (red), connected via microtubules modeled as stiff springs (blue) to the actin cortex, which is modeled as a number of beads connected by weaker springs (green). (b) Cell division. (c) Growing cells at 70\% coverage. Where cell boundaries touch, they coincide with the Voronoi boundaries of their nuclei. (d) Growing cells at 100\% coverage (no division). (e) Growing and dividing cells at 98\% coverage after two divisions. (f) Growing and dividing cells at 98\% coverage after two divisions, for the case in which one initial cell (with four daughters, indicated in orange) has a growth rate that is $2.5\times$ larger than that of the others.}
\label{fig:model}
\end{figure}

We model the cells in two dimensions, treating them as purely mechanical objects. Our cells consist of a nucleus, a radial and stiff microtubule network, and a more flexible actin cortex at the cell perimeter~\citep{Gittes1993}. We model the nucleus as a single large bead with radius $R_\mathrm{n}$, and the cortex as a collection of $M$ small beads with radius $R_\mathrm{c}$ that surround the nucleus (Fig.~\ref{fig:model}a). The cortical beads initially form a circle around the nuclei. We connect each bead to its two neighbors by a spring with spring constant $k_\mathrm{c}$ and rest length $u_\mathrm{c} = 2 R_\mathrm{c}$ to mimic the forces in the actin cortex. Cortical beads that are not connected through these springs interact via the repulsive part of the same potential. Microtubules are modeled as springs that connect the nuclear bead to individual beads in the membrane. To do so, we select at random a fraction $f = 1/6$ of the cortex beads and connect them to the nuclear bead with a spring of spring constant $k_\mathrm{MT}$ and rest length $u_\mathrm{MT} = 2 R_\mathrm{n}$. 

We initiate our system by placing $N$ non-overlapping, circular cells at random positions in the plane. To let the cells grow, we allow the rest length of the microtubules and actin filaments to increase linearly over time. Because cells cannot interpenetrate, they exert forces on each other when they touch. These forces counteract the growth of the microtubules, which halts at a given stall force. A microtubule stops extending when the membrane bead it is connected to comes within 99\% of the minimal equilibrium distance to a bead of another protocell. In this event we also lock the relative position of the beads. When half of the microtubules have stopped growing, the growth of the actin filaments also stops.

To let the cells divide, we first double the number of beads in the membrane and the number of microtubules connecting them to the nucleus. We then split the nucleus into two daughter nuclei of half the size. Of the cortical beads connected to a microtubule, we select the two beads forming the shortest axis across the cell. We then use this axis to divide the microtubules over the two nuclei (Fig.~\ref{fig:model}b). To help the nuclei separate, an extra spring is positioned between the nuclei, mimicking the interpolar microtubules. The rest length of the interpolar spring is gradually increased from zero to the radius of the nucleus, while the rest length of the other microtubules is reduced with a factor~$\sqrt{2}$, so that the total area of the cell remains the same. Once this process is completed, the two axis beads are contracted using a new spring, and when brought together, duplicated and re-connected to complete the division of the cells.

The dynamics of the cytoskeleton and the nuclei are overdamped because the inertia of these small cell components is negligible compared to their viscous drag. Therefore, our equation of motion follows from equating the net force to the drag force, as given by Stokes' law:
\begin{equation}
\label{Stokeslaw}
\mathbf{F}_{i,\mathrm{net}} = 6 \pi \eta R_i \mathbf{v}_i,
\end{equation}
where $\mathbf{F}_{i,\mathrm{net}}$ is the total (net) force on object $i$, which can be either a nuclear or a cortical bead. The viscosity is denoted by $\eta$, $R_i$ is the radius of object~$i$, and $\mathbf{v}_i$ is its velocity.

In our simulations, we scale our measure of length by setting $R_\mathrm{c} = 1$. For the repulsion between two cortical beads we can define a characteristic time $\tau \equiv 6 \pi \eta / k_\mathrm{c}$. We non-dimensionalize the units of time and force by setting $\tau = k_\mathrm{c} = 1$. 

We introduce a quality number $Q$ to quantify the match between the Voronoi tessellation of the nuclei and the actual cells. To do so, we compare the actual area, $A_\mathrm{r}$, of the cells to the area of their corresponding Voronoi cells, $A_\mathrm{V}$. We define $Q$ as:
\begin{equation}
\label{defQ}
Q = \frac{1}{N} \sum_{i=1}^N \left( \frac{A_{\mathrm{r},i} - A_{\mathrm{V},i}}{A_{\mathrm{r},i}} \right)^2,
\end{equation}
where $N$ is the total number of cells. When the Voronoi tessellation has a perfect match with the actual cells the value of $Q$ is 0. For comparison, the $Q$ number for a random close packing of identical discs is 0.05.

\subsection{Experimental system}
To be able to concurrently observe the nuclei and the actin cortex of \textit{D.~melanogaster} and \textit{T.~castaneum}, we required lines in which both parts are fluorescently labeled. For \textit{D.~melanogaster}, we used His2A-RFP/sGMCA flies (Bloomington Drosophila Stock Centre number 59023) that ubiquitously express Histone2A fused to Red Fluorescent Protein (RFP) and the Actin-binding domain of Moesin fused to Green Fluorescent Protein (GFP)~\citep{Kiehart2000}. For \textit{T.~castaneum}, we created a line that ubiquitously expresses LifeAct~\citep{Riedl2008} fused to EGFP~\citep{Benton2013}, thus labeling actin. We further crossed this line to an available nuclear-GFP line~\citep{Sarrazin2012}. We called the crossed line LAN-GFP. Details of the process can be found in appendix~\ref{app:expmethods}.


\section{Results}
First, we observe what happens when we let our model cells grow without division, using random initial placement and double periodic boundary conditions. Because the cortical beads experience drag, the ones that are not connected to a growing microtubule lag behind those that are. When growing cells touch and connect, the forces from the growing microtubules also feed back on the nuclear bead, which shifts position. Fig.~\ref{fig:model}c shows a snapshot of a simulation in which the cells have reached about 70\% coverage of the plane. Where neighboring cells touch, their boundaries coincide with the Voronoi tessellation of the nuclei. The $Q$ number for this case is fairly high ($Q=0.22$), representing the fact that there are still big gaps between the cells. When we let the cells grow further, they eventually reach 100\% coverage, and their geometrical pattern matches the Voronoi tessellation of their nuclei almost perfectly ($Q=2.1\cdot 10^{-3}$, Fig.~\ref{fig:model}d). If we let the cells divide during the developmental process, the final picture is much the same, with again an almost perfect match to the Voronoi tessellation ($Q=3\cdot 10^{-3}$, Fig.~\ref{fig:model}e). However, if we give one of the initial cells a larger growth rate (inherited by its daughters), we find that this pattern is broken (Fig.~\ref{fig:model}f). The faster-growing cells cover a larger fraction of the available area than their corresponding Voronoi cells, whereas their slower-growing neighbors are left with a compressed shape.

\begin{figure}
\begin{center}
\includegraphics{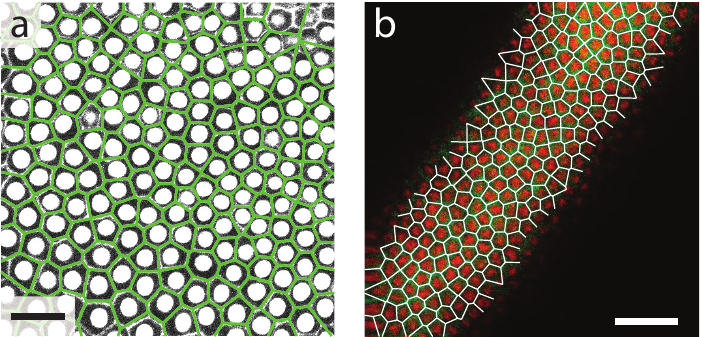}
\end{center}
\caption{Epithelial blastoderm after cellularization of (a) \textit{Tribolium castaneum} and (b) \textit{Drosophila melanogaster} embryos. Overlaid in both images is the Voronoi tessellation of the centroids of the nuclei. Scale bars are~$25\;\mu\mathrm{m}$.}
\label{fig:experiments}
\end{figure}

In both insect systems we studied, the picture is very similar to the simulation results. (Proto)cells appear on the surface at random positions, and grow to confluency after two (\textit{Tribolium}) or three (\textit{Drosophila}) divisions (supplemental figures~\ref{fig:Tribolium} and~\ref{fig:Drosophila} in~\ref{app:observations}). When the cells cover 100\% of the available area, their boundaries also closely match the Voronoi tessellation of their nuclei, as shown in Fig.~\ref{fig:experiments}. Moreover, in both the experimental and simulation results, we find that the positions of the nuclei are close to (though not exactly on top of) the centroids of the Voronoi cells (supplemental figure~\ref{fig:Voronoicentroids}).

To quantify the match between the experimental and the numerical results, we determine the value of two geometrical and one topological property of the cells after cellularization. First, we measure the variance of the area per Voronoi cell, which we find to be very low in both embryos and in the simulations (Table~\ref{tab:geometricmeasures}), indicating that all cells grow to roughly the same size. Second, we measure the reduced area (or circularity) $A^*$ per Voronoi cell, defined as $A^*=4 \pi A/P^2$, where $A$ is the area and $P$ the perimeter of the cell~\citep{Hocevar2009}. With this definition, circles have a reduced area of 1, and hexagons have reduced area of $\pi/2\sqrt{3} \approx 0.91$. We find that the average reduced area of the Voronoi cells in both our experimental systems and in our simulations is again a close match with a value of about 0.83 (Table~\ref{tab:geometricmeasures}).

\begin{table}
\begin{tabular}{|l|c|c|c|c|c|}
\hline
&\multicolumn{2}{c|}{Experiments}&\multicolumn{3}{c|}{Simulations}\\
\hline
&\textit{T. castaneum}&\textit{D. melanogaster}&no division&with division&unequal growth\\
\hline
Area variance & $0.02 \pm 0.02$ & $0.05 \pm 0.005$ & $0.02 \pm 0.005$ & $0.01 \pm 0.002$ & $0.037 \pm 0.004$ \\
Reduced area $A^*$ & $0.85 \pm 0.02$ & $0.83 \pm 0.02$ & $0.83 \pm 0.01$ & $0.83 \pm 0.01$ & $0.83 \pm 0.01$\\
$Q$ number & $0.0009$ & $0.02$ & $0.002$ & $0.003$ & $0.017$ \\
\hline
\end{tabular}
\caption{Values of the two geometrical measures and quality number of the Voronoi tessellations of our experimental and simulated systems. The variance of the area is very small in the first four cases, indicating that in each case, all resulting cells have roughly the same size. In the case where a single cell grows 2.5 times faster than the others (last column), we immediately get a significant increase in this variance. The reduced area (area divided by the perimeter squared normalized such that a circle has a value of 1) is very similar in all cases. Notably, the reduced area is significantly less than that of a regular hexagon (0.91), consistent with the topological observation that only about half of the cells in our system have six vertices.}
\label{tab:geometricmeasures}
\end{table}

\begin{figure}
\begin{center}
\includegraphics[scale=1]{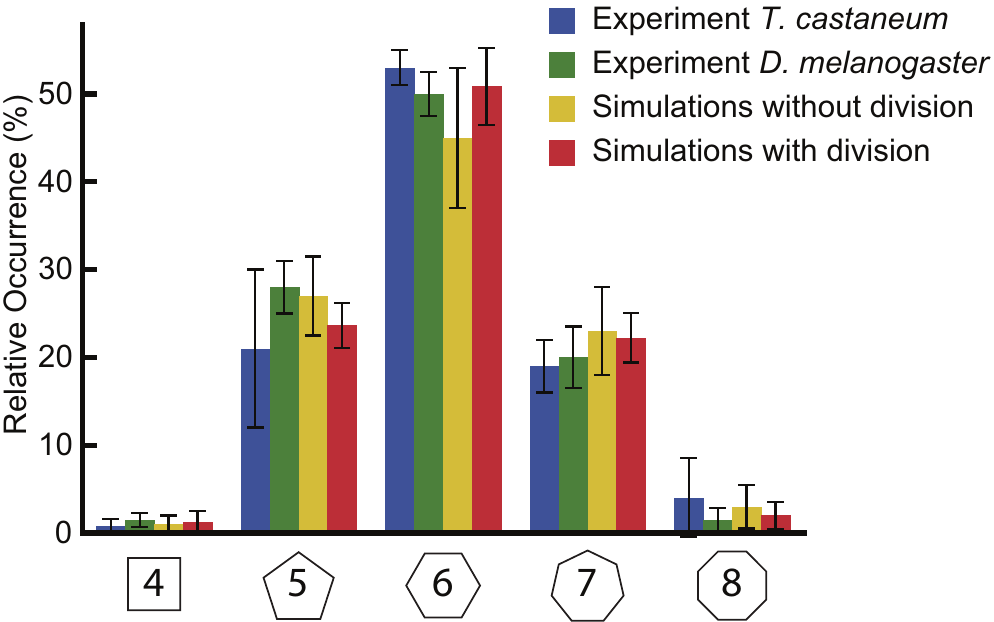}
\end{center}
\caption{Relative occurrence of cells with a given number of vertices in the Voronoi tessellation after cellularization for our experimental observations on \textit{T. castaneum} (blue) and \textit{D. melanogaster} (green) and our simulations without (yellow) and with (red) division. Note that where a perfect honeycomb lattice would consist exclusively of hexagons, only about half of our cells have six vertices, in both experiments and in the simulations.}
\label{fig:vertexfreqdist}
\end{figure}

In addition to the two geometrical measures given above, we also consider a topological measure: the relative occurrence of cells with a given number of vertices. For a perfectly regular pattern (a honeycomb lattice), all cells are hexagons, and thus all cells have six vertices. Deviations from this pattern occur in the form of cells with five and seven vertices (with the total number of vertices of all cells being conserved), or even four or eight vertices. Not surprisingly, hexagonal cells are most abundant in our Voronoi tessellations. However, we also find large numbers of pentagons and heptagons, which each account for about 25\% of the cells (Fig.~\ref{fig:vertexfreqdist}). Again, the two experimental systems and the simulations all agree quantitatively.

\section{Discussion}
Our simulations consistently predict that the growing or growing-and-dividing cells will create a spatial pattern that closely matches the Voronoi tessellation of their nuclei. We observe the same pattern in the first epithelial tissue in our two experimental systems. Moreover, the distribution of cell shapes that we find in the experiments is reproduced exactly by the simulations, without the need for fine-tuning any parameters. Earlier models, which start from a Voronoi tessellation, require a large number of adjustment steps to reach this distribution~\citep{Weliky1990,Farhadifar2007,Staple2010,Sanchez-Gutierrez2016,Barton2017,Lin2017}. Our model instead provides a mechanism for constructing the Voronoi tessellation directly.

In both experimental systems and in the simulations, we measure a reduced area $A^*$ of the cells of about 0.83, just below the order-disorder phase transition reported by Ho\v{c}evar and Ziherl at $A^*=0.865$~\citep{Hocevar2009}. For higher values, epithelial tissues consist almost exclusively of hexagons and are ordered. For values of $A^*$ below the critical value, the tissues are disordered and contain considerable fractions of polygons which are not hexagons, as we observe in our systems. Recent work by Bi et al.~\citep{Bi2015,Bi2016} showed that at almost the same value of the reduced area ($A^*=0.866$), tissues exhibit a rigidity transition. Bi et al. modeled an active tissue using self-propelled Voronoi cells and found that below the reported critical value, the tissues behave fluid-like, whereas for higher values they are solid-like. The tissues in our insect embryos have a reduced area $A^*$ between 0.83 and 0.85, which classifies them as (just) liquid-like. We verified this classification by calculating the pair correlation function of the positions of the nuclei (supplemental figure~\ref{fig:gr}), which is flat, confirming that the nuclei are spatially disordered. This observation is consistent with the stage of development. After cellularization, the embryos undergo a massive shape change, known as gastrulation, in which the mesoderm is formed. Another round of divisions before cellularization would probably push the system over the critical point into a jammed state, which would make gastrulation much more difficult. On the other hand, the cells must be confluent to form a fairly stable tissue. Our observation that the system exists just on the liquid side of the jamming transition may therefore well correspond to a necessary step in development. This `development up to jamming' might also underlie the different number of nuclear divisions before cellularization in different insects~\citep{Anderson1972,VanderZee2015}.

As our simulations show, despite the fact that the cells are placed on the surface randomly, they all reach the same final size (as illustrated by the low variance in the area). We again observe the same effect in the experiments. However, if some cells grow faster than others, we find that the regular pattern is broken. The faster-growing cells end up being larger than the others, and they moreover break the Voronoi tessellation, as their actual boundaries lie well outside their Voronoi cell. These results indicate that the Voronoi patterns observed in many epithelial tissues are due to the mechanical interactions between the proliferating cells that build the tissue, and that those cells must all grow at the same rate.

\section{Conclusion}
We modeled the development of a confluent epithelial tissue from identical cells that are initially distributed randomly. We observe that the resulting configuration of the cells in the tissue closely matches the Voronoi tessellation of their nuclei. We experimentally find the same behavior for the newly formed cells of the epithelial blastoderm in both \textit{D. melanogaster} and \textit{T. castaneum}. We find in both simulations and experiment that in the specific tessellation the cells form, they all have roughly the same area, and the distribution of cell shapes is identical for experiment and simulations. Moreover, the arrangement of the cells is such that the resulting tissue is just on the liquid side of the jamming transition. We can understand the formation of this pattern from mechanical interactions between the cells. Growing cells eventually come into contact with their neighbors, resulting in mechanical feedback that causes them to stop growing towards that neighbor. These contacts moreover translate back to a mechanical force on the nuclei of the cells, which causes them to re-position and eventually form the observed Voronoi tessellation. Thus, mechanical interactions largely determine cell arrangement and shape in epithelial tissues.

\section{Author contributions}
\noindent R.v.D. developed and performed simulations; analyzed data; wrote the manuscript.\\
T.V.-F. performed experiments; analyzed data; wrote the manuscript.\\
T.A.P.M.H. developed and performed simulations; analyzed data.\\
M.v.d.Z. designed research; wrote the manuscript.\\
T.I. designed research; wrote the manuscript.\\
All authors declare no conflict of interest.

\section{Acknowledgments}
T.V.-F. was supported by personal grant nr. 405855 from the Mexican Council for Science and Technology (CONACYT). The funding source had no involvement in the design, execution and reporting of the project. Fly stocks were obtained from the Bloomington Drosophila Stock Centre (NIH P40OD018537).

\appendix
\section{Experimental methods}
\label{app:expmethods}
\subsection{Initial stages of insect development}
In this paper, we study the formation of the first tissue, the epithelial blastoderm, in the embryos of two model organisms: the fruit fly \textit{Drosophila melanogaster} and the flour beetle \textit{Tribolium castaneum}. Like all insects in their early stage of development, these embryos form a syncytium: the nuclei are not separated by plasma membranes into cells, but they are embedded in a full cellular apparatus, compartmentalized by the cytoskeleton. We refer to these compartments as protocells, which thus all share a single cytoplasm. During the first nuclear divisions the nuclei reside in the inside of the egg. After nuclear division 8 (i.e., during cycle 9), the nuclei migrate to the surface of the egg and form a single layer of nuclei, called the syncytial blastoderm. Once the nuclei are at the periphery of the egg, the plasma membrane surrounding the entire egg moves in between the nuclei (invagination) and separates them. Finally, in cycle 13 (i.e., after nuclear division 12) in \textit{T.~castaneum}~\citep{Handel2000}, or cycle 14 (after nuclear division 13) in \textit{D.~melanogaster}~\citep{Foe1983}, the plasma membrane closes around the protocells, creating actual cells. This process, known as cellularization, turns the syncytial blastoderm into a proper epithelial tissue, known as the cellular blastoderm.

\begin{figure}[ht]
\begin{center}
\includegraphics[scale=1]{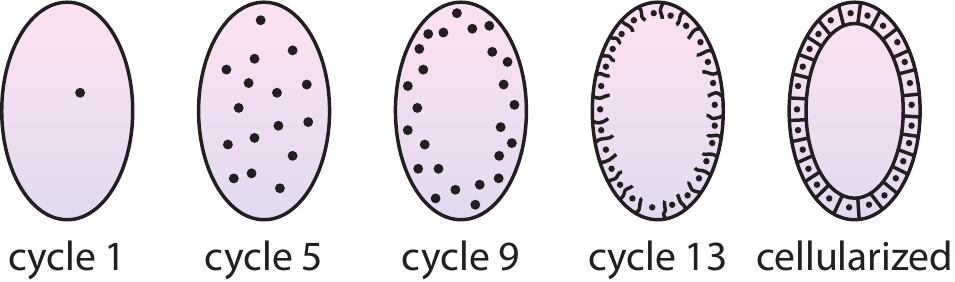}
\end{center}
\caption{Sketch of the syncytial stage of the development of \textit{Tribolium castaneum}. During the first eight cycles, the nuclei duplicate in the bulk of the egg. In the ninth cycle, the nuclei migrate to the surface of the egg, forming the syncytial blastoderm. After four more nuclear divisions, during the thirteenth cycle, cellularization occurs, and the syncytium is converted into a cellular blastoderm tissue.}
\label{fig:syncytium}
\end{figure}

Remarkably, the process of cellularization happens in a different fashion in our two model embryos. In \textit{D.~melanogas\-ter}, actin filaments cover the tips of the invaginating membrane and form a network that might actively be pulled inward by myosin motors. When the membrane furrows have reached the basal side of the protocells, an actin-myosin ring is constructed and contracts to complete cellularization~\citep{Mazumdar2002,Lecuit2004,Harris2009}. In contrast, no such actin-myosin ring is formed in \textit{T.~castaneum}. Instead, the clamping protein Innexin7a forms junctions between the developing basal membrane at the bottom of the furrows and the yolk plasmalemma underneath the protocells. These junctions act as patch-clamps, allowing the basal membrane to spread until it closes off the protocell~\citep{VanderZee2015}. The initial steps of the development of \textit{T.~castaneum} are illustrated in Fig.~\ref{fig:syncytium}. 

\subsection{Transgenic \textit{Tribolium} line with nuclei and Life-actin labeled with GFP (LAN-GFP)}
Our objective was to create a transgenic line of \textit{Tribolium castaneum} in which both LifeActin and the nuclei were labeled with Green Fluorescent Protein (GFP). To create a LifeActin line, we amplified LifeActin~\citep{Riedl2008} fused to EGFP from the pT7-LifeAct-EGFP vector~\citep{Benton2013} by high fidelity PCR, using a forward primer that introduced an FseI restriction site and a reverse primer that introduced an AscI site. We cloned this fragment FseI-AscI under the alpha Tubulin1 promoter~\citep{Siebert2008} into the \textit{piggyBac} vector~\citep{Horn2000} provided by Peter Kitzmann and Gregor Bucher (Georg August Universit\"at, G\"ottingen). Sequences are available on request. The construct containing \textit{vermillion} under the 3xP3 promoter as marker was injected into \textit{vermillion white Tribolium} strain embryos~\citep{Berghammer2009}. For obtaining stable transgenic lines we performed standard crosses~\citep{Berghammer2009}. Two lines in which LifeAct-GFP is ubiquitously expressed all over oogenesis, embryogenesis and larval life were selected. 

To obtain the transgenic line used in this paper, we crossed our stable LifeAct-GFP line to an existing line ubiquitously expressing GFP extended with a Nuclear Localisation Signal (NLS)~\citep{Sarrazin2012}. We named the combined line LAN-GFP. 

\subsection{Maintenance}
\textit{T. casteneum} was kept at $30^\circ\mathrm{C}$ with a humidity of 50\% in a box with wheat flour and dry yeast (1:0.05 w/w) which was refreshed weekly~\citep{Berghammer2009}. Flies were kept under standard conditions on standard medium~\citep{Greenspan1997}.

\subsection{Live imaging}
We put \textit{T.~castaneum} on fine flour at $30^\circ\mathrm{C}$ for one hour. Subsequently, we removed the adults using a sieve with a $600-850\;\mu\mathrm{m}$ mesh size and collected the eggs using a sieve with a $250\;\mu\mathrm{m}$ mesh size. We then let the eggs develop for four hours at $30^\circ\mathrm{C}$. After the four hours, we dechorionated the eggs in 5\% bleach. We put \textit{D.~melanogaster} on egg laying at $25^\circ\mathrm{C}$ for one hour. We subsequently collected the eggs and immediately dechorionated them with 5\% bleach. We lined the eggs of both insects on a microscope glass-bottomed Petri dish (Willco Wells BV). To avoid desiccation of the eggs we covered them with Voltalef 10S Halocarbon oil.
 
\begin{figure}[ht]
\begin{center}
\includegraphics[scale=1]{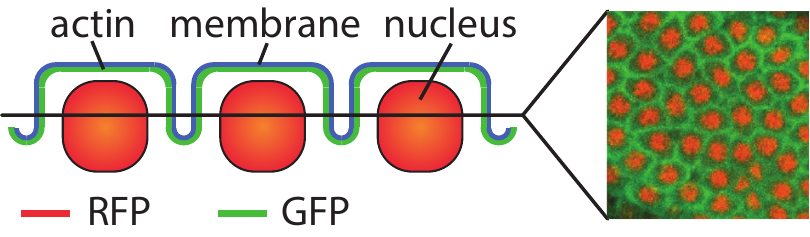}
\end{center}
\caption{Example of a summed z-stack (11 images) of the syncytial blastoderm of \textit{D.~melanogaster} (right), with histones labeled with RFP (red) and actin with GFP (green). As shown in the schematic on the left, although the two-dimensional cross section of the protocells shows no holes, at this stage (before cellularization) the plasma membrane does not yet separate these protocells into complete cells.}
\label{fig:membranecrosssection}
\end{figure}

We imaged the embryos on an inverted Zeiss confocal microscope at $30^\circ\mathrm{C}$ for \textit{T.~castaneum} and $25^\circ\mathrm{C}$ for \textit{D.~melanogaster}. We observed the embryos at a cross-section of the syncytial blastoderm (Fig.~\ref{fig:membranecrosssection}). We took z-stacks consisting of eleven focal planes with a $40\times/1.5\;\mathrm{N.A.}$ water-immersion objective. The time interval between frames was 3~minutes for \textit{T.~castaneum} and 1~minute for \textit{D.~melanogaster}. The total observation time was six hours for \textit{T.~castaneum} and two hours for \textit{D.~melanogaster}. 

We used imageJ 1.49t to process the images and to obtain the area of the cells, using the method developed by Aigouy~\citep{Farhadifar2007}. We summed all the planes in the z-stack at each time point to make a time-lapse video. 

\subsection{Acknowledgments}
We thank Peter Kitzmann and Gregor Bucher (Georg August Universit\"at G\"ottingen, Germany) for help with generating the transgenic \textit{Tribolium} line; Matthew Benton (University of Cologne, Germany) for providing the pT7-LifeAct-EGFP construct and for help with live imaging; Ron Habets (Leiden University Medical Centre, The Netherlands) for help with selecting Drosophila stocks; Gerda Lamers (Leiden University, The Netherlands) for help with confocal microscopy; Kees Koops and Onno Schaap (Leiden University, The Netherlands) for taking care of the beetles and flies.

\section{Observations}
\label{app:observations}
\subsection{Experiments}
\begin{figure}[ht]
\begin{center}
\includegraphics[scale=1]{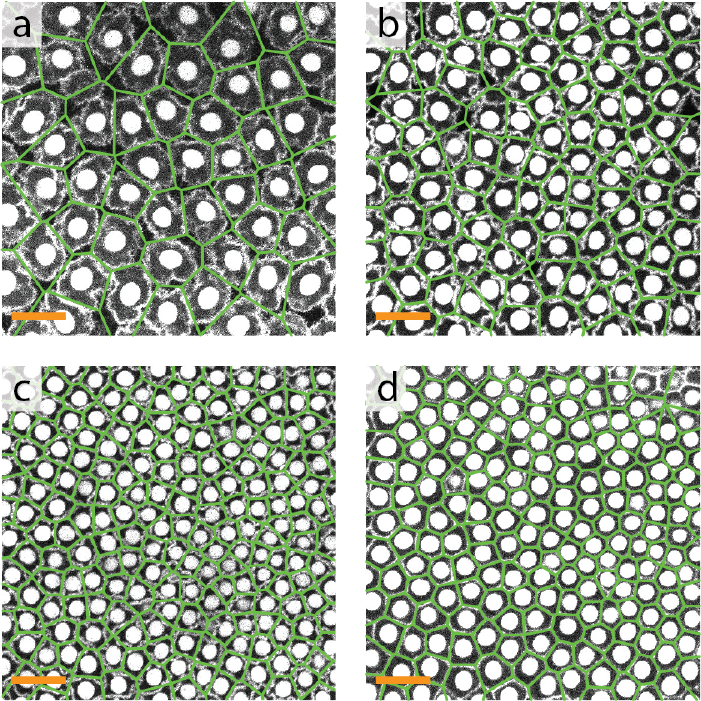}
\end{center}
\caption{Four snapshots from a video of a \textit{Tribolium castaneum} embryo from our new LAN-GFP line, in which both the actin and the nuclei are labeled with GFP. Overlaid in green is the Voronoi tessellation of the centers of the nuclei. Figures a-c show snapshots after the 10th, 11th and 12th nuclear division before cellularization, and figure d shows a snapshot of the embryo after cellularization. Scale bar $25\;\mu\mathrm{m}$.}
\label{fig:Tribolium}
\end{figure}

We used our new \textit{Tribolium} line expressing LifeAct-EGFP and nuclearGFP (LAN-GFP) to concurrently visualize the cortical actin and the nuclei. Using the existing line His2A-RFP/sGMCA, we similarly visualized the cytoskeleton and nuclei in \textit{D.~melanogaster}. For each of the images, we determined the position of the centers of all the nuclei. From these, we constructed the associated Voronoi tessellation (Fig.~\ref{fig:Tribolium}a-d), i.e., the division of space into cells such that each point is part of the cell corresponding to the closest center. We overlaid these Voronoi tessellations with the experimental images for the last three cycles in the syncytium, and after cellularization.

Although the protocells are basally connected by the plasma membrane of the entire egg, the lateral sides of the protocells are not tightly adjacent and leave space between the protocells in \textit{T.~castaneum} (Figs.~\ref{fig:Tribolium}a and b) until the last cycle before cellularization (Fig.~\ref{fig:Tribolium}c). In the 11th cycle, the protocells only cover about 70\% of the available space, rising to 85\% in cycle 12 and close to 100\% in cycle 13 before cellularization. Consequently, the Voronoi tessellation of the nuclei does not correspond closely with the position of the protocellular boundaries in cycles 11 and 12 (Figs.~\ref{fig:Tribolium}a and b). The quality numbers are $Q=0.16$ and $Q=0.12$, respectively. However, we note that where two adjacent protocells touch, their boundary typically does follow the boundary between the corresponding Voronoi cells. In cycle 13 before cellularization, the match between the Voronoi tessellation and the protocellular boundaries is better (Fig.~\ref{fig:Tribolium}c), although the quality number, $Q=0.10$ still indicates a fairly poor match. The reason for this poor match is that, unlike the tessellation boundaries, the cellular boundaries are not straight at this point, indicating that they are not under tension. After cellularization, when the boundaries of the newly formed cells are under high tension and the membrane is straight, the match between the experimental data and overlaid tessellation becomes almost perfect (Fig.~\ref{fig:Tribolium}d) with a $Q$ number close to zero ($Q = 9\cdot 10^{−4}$). 

\begin{figure}[ht]
\begin{center}
\includegraphics[scale=1]{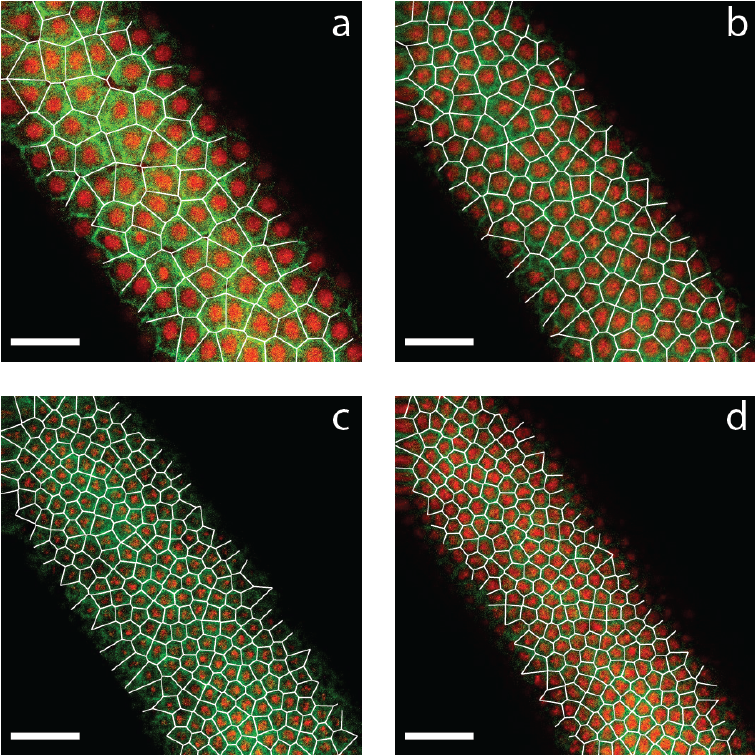}
\end{center}
\caption{Four snapshots from a video of a \textit{Drosophila melanogaster} embryo in which the nuclear histones are labeled with RFP (red) and the actin with GFP (green). Overlaid in white is the Voronoi tessellation of the centers of the nuclei. Figures a-c show snapshots after the 11th, 12th and 13th nuclear division before cellularization, and figure d shows a snapshot of the embryo after cellularization. Scale bar $25\;\mu\mathrm{m}$.}
\label{fig:Drosophila}
\end{figure}

The picture for \textit{D.~melanogaster} is largely similar to that of \textit{T.~castaneum}, though different in the details. The nuclei of \textit{D.~melanogaster} divide once more before cellularization. After nuclear division 11, the cells cover about 75\% of the available space, and the Voronoi tessellation has a poor match, with $Q=0.14$, like in \textit{T.~castaneum} (Fig.~\ref{fig:Drosophila}a). The match with the Voronoi tessellation improves after nuclear division 12, with $Q=0.05$ (Fig.~\ref{fig:Drosophila}b). In the 14th cycle (after the nuclear division 13), before and after cellularization, we obtain $Q=0.07$ and $Q=0.01$, respectively (Figs.~\ref{fig:Drosophila}c and \ref{fig:Drosophila}d). The pattern of the protocells in \textit{D.~melanogaster} thus already matches the Voronoi tessellation at an earlier stage than happens in \textit{T.~castaneum}. The difference between the two species may be due to two effects. First, as the nuclei in \textit{D.~melanogaster} undergo an additional nuclear division before cellularization, they are more densely packed than those in \textit{T.~castaneum}, resulting in less unoccupied space and hence more mechanical contacts between the protocells. Our simulation results (detailed below) suggest that such mechanical contacts will inevitably lead to the formation of boundaries that correspond to the Voronoi tessellation of the nuclei. Second, as the mechanism of basal cell closure is quite different in both species (active constriction for \textit{D.~melanogaster} versus passive patch-clamp constriction for \textit{T.~castaneum}), the tension in the membrane after cellularization may be different as well.

\subsection{Simulations}
\begin{figure}[ht]
\begin{center}
\includegraphics[scale=1]{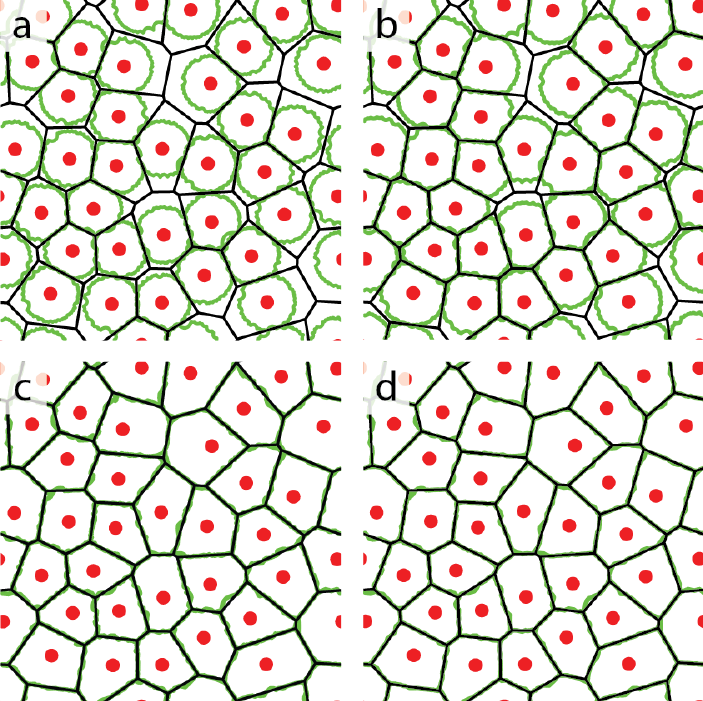}
\end{center}
\caption{Snapshots of the simulation (growth only) with the nuclei depicted in red, the actin cortex in green, and the overlaid Voronoi tessellation of the nuclei in black. Microtubules not shown. Figures a-c show snapshots in which the (proto)cells cover (a) 70\%, (b) 85\% and (c) 100\% of the available area. d) Pattern of the cells after the links in the actin cortex have been put under tension, mimicking the tension on the membrane / actin cortex during cellularization.}
\label{fig:simulations}
\end{figure}

In Fig.~\ref{fig:simulations}, we show the configuration of the membrane and the Voronoi tessellation of the nuclei for different area coverage fractions (Figs.~\ref{fig:simulations}a-c), and after we stretch the membrane (corresponding to cellularization, Fig.~\ref{fig:simulations}d). The coverage percentages of 75\%, 85\%, and close to 100\% shown in Figs.~\ref{fig:simulations}a-c correspond to those we found experimentally in the last three cycles of \textit{T.~castaneum} embryos. In agreement with the experimental data, the Voronoi cell boundaries are a good match for the actual cell boundaries where they touch, but the boundaries of the actual cells do fluctuate instead of forming a straight line. Due to these fluctuations, the match between the Voronoi tessellation and the actual cells is not perfect. Before cellularization we find a match of $Q=0.22$ at 75\% coverage, $Q=0.10$ at 85\% coverage, and $Q=2.1\cdot 10^{−3}$ at 100\% coverage. After cellularization, when the membrane is stretched, the fluctuations around the Voronoi cell perimeter are reduced and the quality number improves ever further to $Q=8.0\cdot 10^{−4}$. An overview of all the obtained quality numbers from the experiments and simulations is given in Table~\ref{table:Qnumbers}.

\begin{table}[ht]
\begin{center}
\begin{tabularx}{\textwidth}{|L{1}|C{1}|C{1}|C{1}|C{1}|}
\hline
$Q$ numbers & After antepenultimate nuclear division & After penultimate nuclear division & After ultimate nuclear division & After cellularization \\
\hline
\textit{D.~melanogaster} & 0.14 & 0.05 & 0.07 & 0.02\\
\textit{T.~castaneum} & 0.16 & 0.12 & 0.10 & 0.0009\\
Simulations & 0.22 & 0.10 & 0.002 & 0.0008\\
\hline
\end{tabularx}
\end{center}
\caption{Quality numbers $Q$ of the match between the Voronoi tessellation and the actual cell boundaries, for both experimental systems and our simulations. For the experimental systems, we calculate the $Q$ number after the last three nuclear divisions and after cellularization, for the simulations we consider the cases shown in Fig.~\ref{fig:simulations}, in which the cell coverage equals that of the \textit{T.~castaneum} embryo in the corresponding nuclear divisions.}
\label{table:Qnumbers}
\end{table}

\subsection{Centroids of the Voronoi tessellations}
\begin{figure}[ht]
\begin{center}
\includegraphics[scale=1]{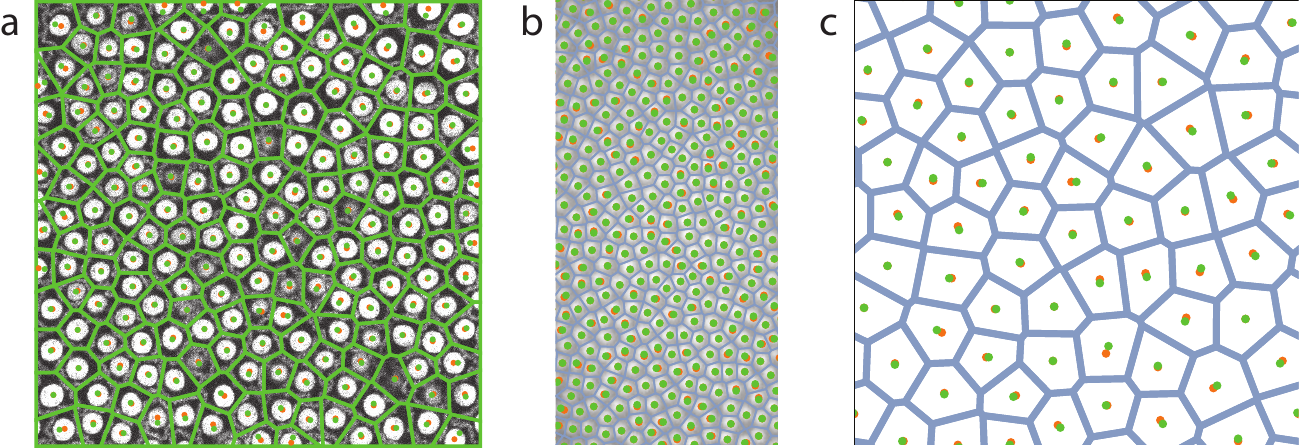}
\end{center}
\caption{Voronoi tessellations of the nuclei of (a) a \emph{Tribolium} embryo, (b) a \emph{Drosophila} embryo, and (c) a simulation at close to 100\% coverage. In all three cases, the positions of the nuclei / organizing centers are indicated with orange dots, and those of the centroids of the Voronoi cells with green dots.}
\label{fig:Voronoicentroids}
\end{figure}

Figure~\ref{fig:Voronoicentroids} shows a typical example of a tessellation of a \emph{Tribolium} and a \emph{Drosophila} embryo and a simulation indicating both the nuclear positions / organizing centers and the centroids.

\subsection{Pair correlation functions}
\begin{figure}[ht]
\begin{center}
\includegraphics[scale=1]{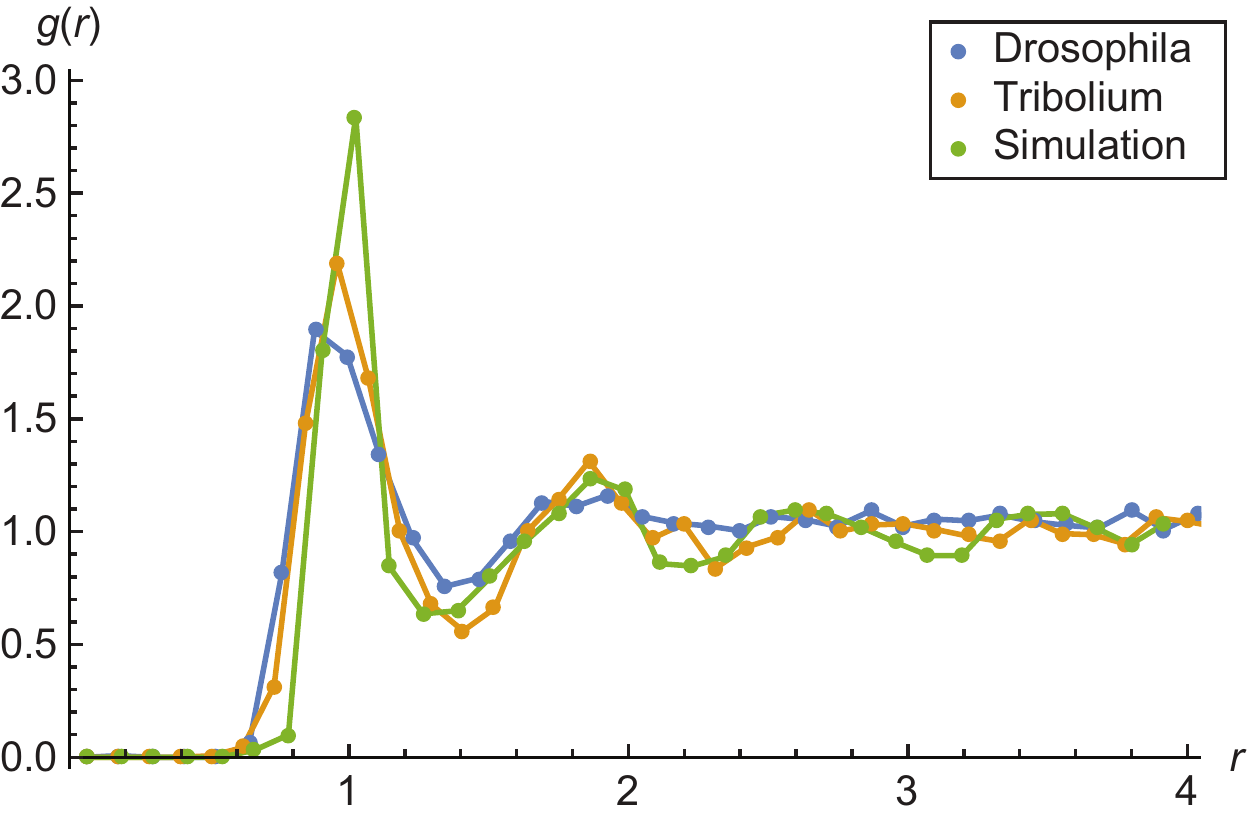}
\end{center}
\caption{Spatial correlation function~$g(r)$ for experimental and simulation data, averaged over five configurations. Distances are normalized by the average separation between two nuclei. The flatness of the correlation function after the initial nearest-neighbor peak indicates that the positions of the nuclei remain randomly distributed, even when the tissue has become confluent, and the system is therefore not ordered. The difference in height of the first peak is due to the fact that there is a distribution in nuclear sizes in the experimental data, while the simulations all nuclei have the same size. The total area under the first peak is the same for both experiment and simulation.}
\label{fig:gr}
\end{figure}

To verify that the nuclei are indeed on the liquid side of jamming, we calculated the pair correlation function~$g(r)$~\cite{ChaikinLubensky} of the nuclei, for both the experimental and simulation data. The results, plotted in figure~\ref{fig:gr}, indicate that the positions are completely uncorrelated (as the $g(r)$ is flat after the initial nearest-neighbors peak), so the nuclei are indeed still in the unordered (liquid-like) phase.

\newpage
\section*{References}

\end{document}